\documentclass[11pt]{article}

\usepackage{graphicx}
\usepackage{caption}
\usepackage{placeins}
\begin{document}
\title{Measurement of T-odd asymmetry in radiative  \boldmath $K^+ \rightarrow \pi^{0} e^{+} \nu_{e} \gamma $  
decay using OKA detector} 



\maketitle

\author{A.~Yu.~Polyarush $^a$\footnote[1]{e-mail: polyarush@inr.ru},
 S.~A.~Akimenko$^b$, A.~V.~Artamonov$^b$, V.~N.~Bychkov$^c$,  S.~V.~Donskov$^b$,
V.~A.~Duk$^{a}$\footnote[2]{Now at INFN Sezione di Perugia, Via A. Pascoli, 06123 Perugia, Italy.}, 
 A.~V.~Inyakin$^b$,  A.~M.~Gorin$^b$,  E.~N.~Gushchin$^{a}$,
 A.~P.~Filin$^b$, S.~N.~Filippov$^{a}$,
G.~D.~Kekelidze$^c$,  V.~N.~Kolosov$^b$, V.~I.~Kravtsov$^{a}$ , 
G.~V.~Khaustov$^b$, S.~A.~Kholodenko$^b$, A.~A.~Khudyakov$^{a}$,
Yu.~G.~Kudenko $^{a}$\footnote[3]{Also at National Research Nuclear University (MEPhI), Moscow, 
and Institute of Physics and Technology, Moscow, Russia},
A.V. Kulik$^a$, V.~F.~Kurshetsov$^b$, 
V.~A.~Lishin$^b$, V.~M.~Lysan$^c$, M.~V.~Medynsky$^b$, 
V.~F.~Obraztsov$^b$, 
A.~V.~Ohotnikov$^b$
V.~A.~Polyakov$^b$, V.~I.~Romanovsky$^b$, V.~I.~Rykalin$^b$,
A.~S.~Sadovsky$^b$, V.~D.~Samoilenko$^b$, 
I.~S.~Tyurin$^b$,
 O.~G.~Tchikilev$^b$,
V.~A.~Uvarov$^b$,
 O.~P.~Yushchenko$^b$,  
B.~Zh.~Zalikhanov$^c$}
\vspace {5pt}


$^a$  Institute for Nuclear Research RAS, 117312 Moscow, Russia.\\
$^b$Institute for High Energy Physics, National Research Center Kurchatov Institute,
Protvino, Moscow region, 142281 Russia\\
$^c$Joint Institute for Nuclear Research, Dubna, Moscow region, 141980 Russia

\begin{abstract}The paper presents a measurement of the T-odd correlation in radiation
decay
\boldmath $K^+ \rightarrow \pi^{0} e^{+} \nu_{e} \gamma $  performed
on the installation of the
101200 candidate events of the investigated decay were identified.
Measured correlation $\xi_{\pi e \gamma}$
-is a mixed product of moments $e^{+}$, $\pi^{0}$, $\gamma$
  in the kaon rest system, normalized by $M^{3}_{K}$.
To assess the asymmetry of the distribution by $\xi_{\pi e \gamma}$
the value used is $A_{\xi}= \frac {N_{+}-N_{-}} {N_{+}+N_{-}}$,
where $N_{+(-)}$ is the number of events with $\xi$ greater than (less than) zero.
For the $A_{\xi}$ asymmetry, the value is obtained
$A_{\xi} = (+0.1 \pm 3.9($stat.$) \pm1.7($syst.$))\times10^{-3}$
or $|A_{\xi}| < 5.4\times10^{-3} (90\%~ CL)$
\end{abstract}
\maketitle
\section{Introduction}
In this article we continue the experimental study
of the decay  $K^+ \rightarrow\pi^{0} e^{+}\nu_{e}\gamma$ ($K_{e 3g}$) started in \cite{e2},
 performed on triple statistics. 
This decay is of great interest because it
allows a search for the $T$-odd triple correlations, by the $CPT$ theorem
 equivalent to detection   of $CP$-invariance violation, which in kaon physics has
been observed so far only in decays of neutral kaons.
Therefore, radiation decays of charged K-mesons are of great interest to
both theorists and experimenters as a possible
alternative source of information about $CP$-invariance violation.

\begin{figure}[ht]
\vspace{-1.5cm}
\mbox{
\includegraphics[width=0.5\linewidth]{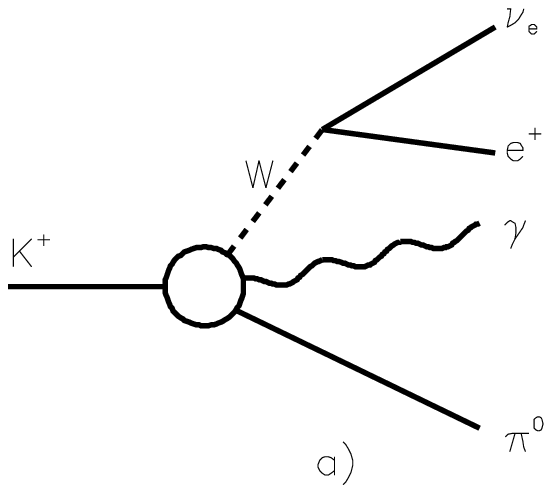}
\hspace{-2.0cm}
\includegraphics[width=0.5\linewidth]{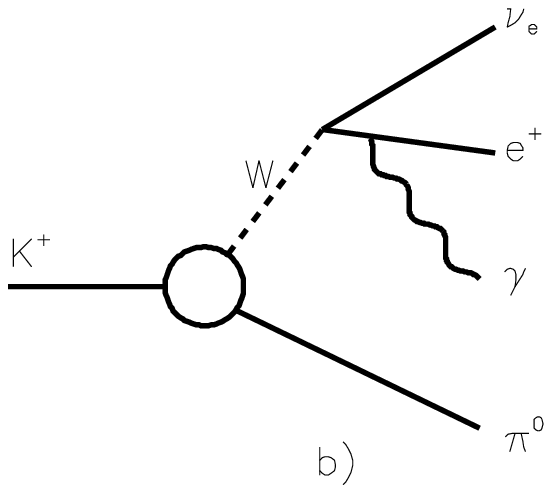}
}
\vspace{-1.5cm}
  \caption{\em Diagram of $K^+ \rightarrow \pi^{0} e^{+} \nu \gamma $ decay }
\end{figure} 

The experimental data currently available on
the violation of $CP$-invariance are explained by the complex phase
of the Kabibbo-Kobayashi-Maskawa quark mixing matrix\cite{v1,v2,v3}.
However, it has been proven that this mechanism is insufficient for
explanations of the observed baryon asymmetry of the universe\cite{v4,v42,v43}.
This forces us to look for new sources of violation of $CP$-invariance.
In general, a search for new processes with violation of $CP$-invariance makes it possible
to look for a new physics beyond the Standard Model (SM).

Although CP-violation is extremely small in SM in leptonic and semileptonic decays
of kaons,
relatively large CP-violating effects are predicted in various models beyond the SM.
So, in the decay
$K^+\rightarrow\mu^{+}\nu_{e}\gamma$ various models\cite{v6, v61,v62, v7,v71,v8}
predict the $T$-odd transverse polarization of the muon at the level from
$5 \times 10^{-3}$\cite{v7,v71} to $5 \times 10^{-2}$\cite{v8},
and in the decay  $K^+ \rightarrow \mu^{+}\nu_{\mu} \pi^{0} $ from
$5 \times 10^{-5}$\cite{v9} to $5 \times 10^{-3}$\cite{v10}.
The experimental restrictions $(90\%~CL)$ for the transverse polarisation are : $P_T < 3.1\times 10^{-2}$ and
$P_T < 5.0\times 10^{-3}$\cite{v11, v111}.
These experiments provide the best opportunities for
detecting the scalar (pseudoscalar) version of the New Physics (NP), while
as it was noted in \cite{tr2}, the decay of $K_{e 3g}$ allows to search for
the vector (axial) variant of the NP for which the matrix element
of the decay
$K^+ \rightarrow \pi^{0} e^{+} \nu \gamma $ has a form:

\begin{eqnarray}
T =\frac{G_{F}}{\sqrt{2}}{e}V_{us}\varepsilon^{\mu} 
(q)\Biggl\{(V_{\mu\nu}
- A_{\mu\nu})\overline{u}(p_{\nu})\gamma^{\nu}(1 - \gamma_{5})v(p_{l})\nonumber\\
+\frac{F_{\nu}}{2p_{l}q}\overline{u}(p_{\nu})\gamma^{\nu}(1 - 
\gamma_{5})(m_{l}-\hat{{p}_{l}}-\hat{q})\gamma_{\mu}v(p_{l})\Biggr\},
\end{eqnarray}
\noindent
 where hadron tensors $V^{had}_{\mu\nu}$ and $A^{had}_{\mu\nu}$ are defined as

$I_{\mu\nu} = i \int d^4 e^{iqx}\langle \pi^{0}(p')|TV^{em}_{\mu}(x)I^{had}_{\nu}(0)|K^+(p) \rangle$, 
$I = V, A,$ with
$V^{had}_{\nu} = (1+g_{V}) \overline s \gamma_{\nu} u$, 
$A^{had}_{\nu} = (1-g_{A}) \overline s \gamma_{\nu} \gamma_{5} u$,
$V^{em}_{\mu} = (2\overline u    \gamma_{\mu} u -  \overline d \gamma_{\mu} d - \overline s \gamma_{\mu} s )/3$
and $F_{\nu}$ is the $K^+_{e3}$ matrix element
$F_{\nu} = (1+g_{V})\langle \pi^{0}(p')|V^{had}_{\nu}(0)|K^+(p) \rangle$
\noindent
here $g_{V}, g_{A}$ are the vector and pseudovector constants, which can be complex.
The first term of equation (1) describes the kaon bramsstrahlung and the
structural radiation diagram Fig.1{\it a}. The lepton bramsstrahlung radiation is represented
by the second part of equation (1) and the diagram Fig.1{\it b}.

For the first time, a search for the triple $T$-odd correlations 
in the radiative decays of K-mesons was proposed in \cite{v12}.
To study the triple $T$-odd correlations, a variable is used

\begin{equation}
\xi_{\pi e \gamma} = \frac{1} {M^{3}_{K}} \vec{p}_{\gamma}\cdot[\vec{p}_{\pi}\times \vec{p}_{l}]
\end{equation}

To estimate the asymmetry of the distribution over the $\xi$ variable, we use the value
\begin{equation}
A_{\xi}= \frac {N_{+}-N_{-}} {N_{+}+N_{-}}
\end{equation}
\noindent
where $N_{+(-)}$ is the number of events with $\xi$ greater than (less than) zero. 
In  paper \cite{tr2} the following theoretical constraint
for the vector and axial versions of the New physics in the framework of a model
based on a gauge group $SU(2)_{L}\times SU(2)_{R}\times U(1)$,
 was obtained:
\begin{equation}
|A_{\xi}(K^+\rightarrow\pi^{0} e^{-}\nu_{e} \gamma)|<0.8\cdot10^{-4} 
\end{equation}

In SM in the tree level, the asymmetry is zero, but a comparable value
of $A_\xi$ appears as a result of electromagnetic
interaction in the final state. This effect
in one-loop approximation was calculated in \cite{v5,v13}, the result is:
$A_\xi = -0.59\cdot 10^{-4}$, and $A_\xi = - 0.93\cdot 10^{-4}$
respectively.

\section{Experiment}
\begin{figure}[ht]
\includegraphics[width=14.0cm]{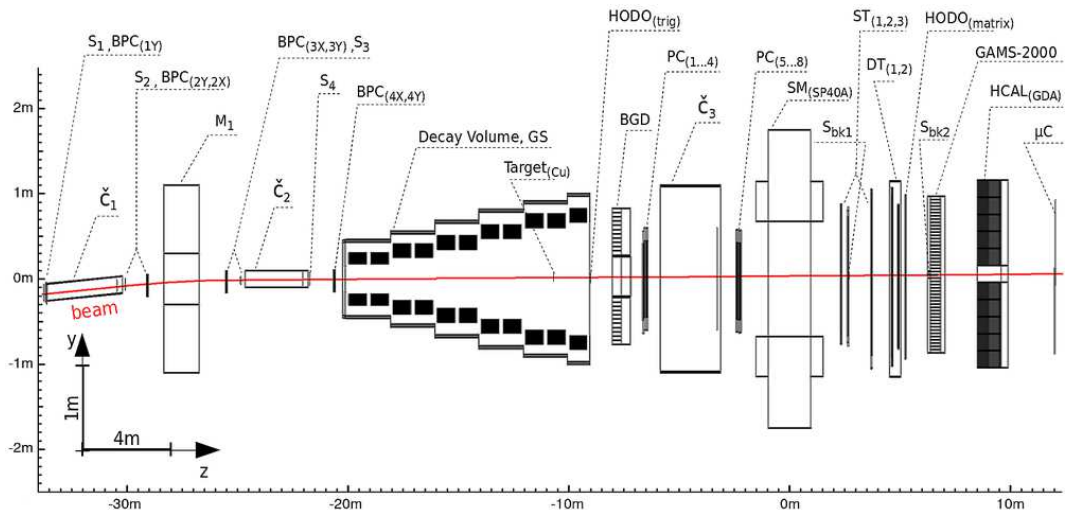}
 \caption{\em Scheme of the OKA setup.}

  \label{figure:ust}
\end{figure}

The OKA experiment has been carried out on the IHEP U-70 proton synchrotron 
on a secondary separated beam of K mesons
with a momentum of 17.7GeV/c, enriched with K mesons up to 20\%. 
The OKA setup is described in detail in our recent publications\cite{e2,e3,e4}
It consists (see figure~\ref{figure:ust}) of a beam spectrometer, a decay volume with a veto system,
a charged particle spectrometer, an electromagnetic calorimeter,
a hadron calorimeter and a muon detector.
The trigger used is described in \cite{e2}.

Monte Carlo (MC) calculations for the background and signal processes was carried
out using the GEANT3\cite{e9} package. Events are weighted according to theoretical
matrix elements.
The signal MC uses the $O(p^{4})$ approximation  of the
Chiral perturbation theory(ChPT)\cite{v5}.

\section{Event selection}

The events with one positively  charged track registered
by the detector's track system and four showers in the GAMS-2000 electromagnetic
calorimeter are selected as the candidates for the $K^{+} \rightarrow \pi^{0} e^{+} \nu_{e}\gamma $ decay.

One of the showers should be associated
with the charged track. The positron identification 
is done using the ratio   of the shower  energy
and the momentum of the positron measured by the tracking system.
In addition, a restriction on the distance
between the charged track extrapolation
to the front plane of the electromagnetic detector and
the nearest shower is used: d$ < $3 cm.

\begin{figure}[!hb]
\vspace{-1.0cm}
\hspace{1.0cm}
\includegraphics[height=9cm]{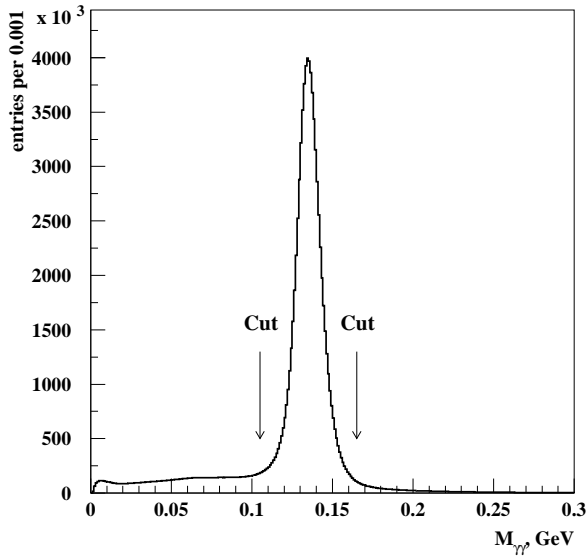}
\caption{\em The invariant mass off $\gamma\gamma$-pairs 
closest to the table value of  $\pi^{0}$ mass}
\label{figure:piz2}
\end{figure}

To reconstruct $\pi^{0}$, a pair from the three remaining
showers (photons) not associated with the track with an invariant mass
closest to the table value of the mass of $\pi^{0}$ is used (Fig. ~\ref{figure:piz2}).
To suppress the background, a selection $|m_{\gamma \gamma}$ - $m_{\pi^{0}}| < $30 MeV is used.
The energy of the photons included in $\pi^{0}$ should be greater than 0.5 GeV.
The energy of the remaining photon should exceed 0.7 GeV.

\section{Background suppression}

The main background decay channels for the investigated decay are:

1) $K^+\rightarrow\pi^{+}\pi^{0}\pi^{0}$ with 1 lost $\gamma$
and $\pi^{+}$ falsely identified as a positron.

2) $K^+\rightarrow\pi^{+}\pi^{0}$ with a random additional $\gamma$ 
and $\pi^{+}$ falsely identified as a positron.

3) $K^+\rightarrow \pi^{0} e^{+}\nu_{e}$ with additional
$\gamma$  due to the interaction of $e^{+}$ with the set-up substance.

4) $K^+\rightarrow\pi^{+}\pi^{0}\gamma$
with $\pi^{+}$ falsely identified as a positron.

5) $K^+ \rightarrow \pi^{0} \pi^{0} e^{+} \nu_{e} $ with 1
$\gamma$ lost.
All these background processes are included in the Monte Carlo calculations.

To suppress the backgrounds (1) - (5), we use selections:

Cut 1: $E_{miss} = E_{beam} - E_{detected} >0.5 GeV$.

The requirement on the missing energy mainly reduces the background (4).

Cut 2: $\Delta y = | y_{\gamma} - y_{e^{+}} | >$ 3 cm, where y is the vertical coordinate
of the point where the positron and photon hit the electromagnetic calorimeter.
The magnetic field rotates the track of the charged particle in the xz plane.
This selection suppresses, first of all, the background (3) from 
the decay of $K^+\rightarrow\pi^{0} e^{+}\nu_{e}$.

Cut 3: $\mid x_{\nu},y_{\nu} \mid <$100 cm.
A straight line along the direction of the missing momentum
 must cross the aperture of the electromagnetic calorimeter.
This selection helps to suppress the backgrounds (1,5) where there are lost photons.

Cut 4: $0.004 < \Theta_{e\gamma}$ $ <0.080$ rad.
The left part of this selection is introduced precisely to suppress the background (3).
The right part of the selection is applied against the background (2) of $K_{\pi2}$ decays.

\begin{figure}[ht]
\hspace{1.0cm}
\includegraphics[width=9.0cm]{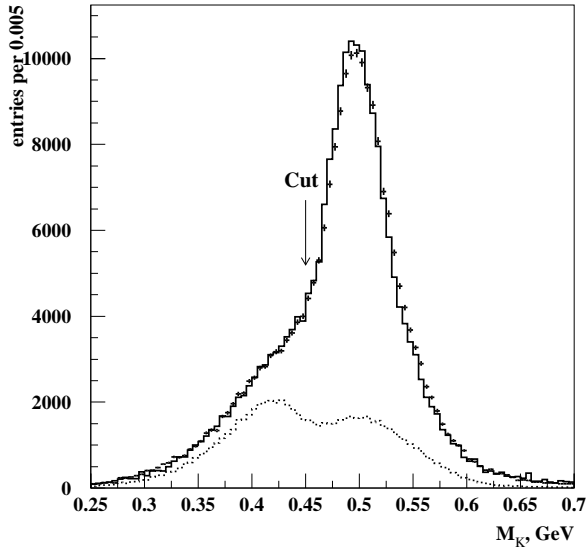}
\caption{\em Distribution over the reconstructed kaon mass.
The dotted curve is the total background.
A solid histogram is the sum of the MC  signal and background.}

\label{figure:piz9}
\end{figure}

Cut 5: $M_{K\rightarrow\pi^{0}e^{+}\nu_{e}\gamma}> 0.45$GeV.
$M_{K\rightarrow \pi^{0}e^{+}\nu_{e}\gamma} $ - the reconstructed mass of the 
($\pi^{0} e^{+} \nu_{e}\gamma$) systems, assuming
that the mass of an unregistered particle is zero ($m_{\nu}=0$).
The distribution of $M_{K}$ at this stage of the selections is shown in Fig.~\ref{figure:piz9}.

Cut 6: $-0.006 < M^{2}_\nu <0.006$GeV$^2$.
\noindent
To strengthen the selection  5, we use the requirement for the missing mass squared
$M^{2}_\nu = (P_{K} - P_{\pi^{0}} - P_{e} - P_{\gamma})^2$.
For signal events, this
variable corresponds to the square of the neutrino mass
and must be zero within the measurement accuracy,
and for background events, the distribution for this variable is much wider.

The dominant background for the    $K_{e3\gamma}$ decay
is that from $K_{e3}$ decay with an additional photon - background (3).
This background is suppressed by selection 2, as well as by the cut on
the angle between the positron and the photon in the laboratory system
$\Theta_{e \gamma}$.
The distribution of $K_{e3}$-background events has a very narrow peak at zero $\Theta_{e \gamma}$.
This peak is much narrower than in signal events.
This is because the emission of photons by the positron in the background process
occurs as a result of $e^{+}$ interactions in the detector material after the decay vertex,
and the angle in the reconstruction program is still calculated,
as if the radiation was emitted from the primary vertex.

The background decay channel (4) has a branching at the level of the studied one and
is suppressed by the correct identification of the positron,
as well as the missing energy selection(1).

The background channel (5) is suppressed by the selection of the missing mass (Cut 6).

As a result, after all the selections, we are left with 101200 candidate events for the
decay of $K^{+} \rightarrow \pi^{0} e^{+}\nu_{e}\gamma$.
The background  is 17700 events.
The normalization of backgrounds was carried out by comparing
the number of registered $K^+\rightarrow\pi^{0} e^{+}\nu_{e}$ decays
in the data and MC.

\section{Results}

\begin{figure}[ht]
\hspace{1.0cm}
\includegraphics[width=9.0cm]{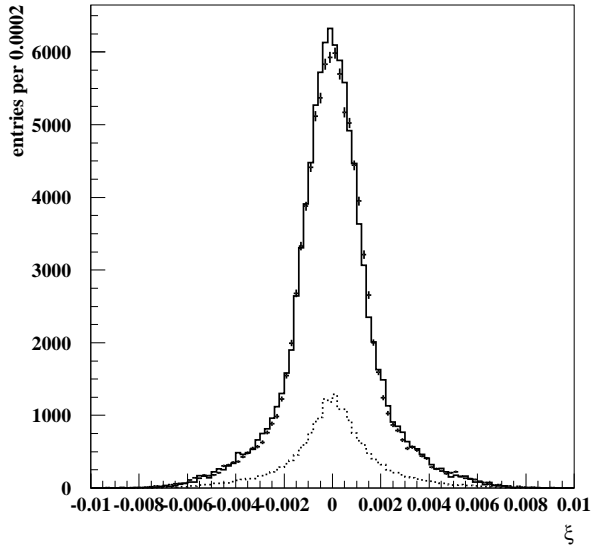}
\caption{\em Distribution over  $\xi$ variable.
The dotted curve is the total background.
The solid histogram is the sum of the MC signal and background.}
\label{figure:piz19}
\end{figure}

Violation of T-invariance leads to asymmetry in the distribution over
the variable $\xi$  (2), shown in Fig. ~\ref{figure:piz19}.

The measured value of $A_{\xi}$ (3), characterizing the asymmetry,
is calculated 
for $E^{*}_{\gamma}> 10$ MeV and $\Theta^{*}_{e\gamma}> 10^{\circ}$,
where $E^{*}_{\gamma}$, $\Theta^{*}_{e\gamma}$ - the photon energy and emition angle in the kaon rest frame

$A_{\xi} =(+0.1 \pm 3.9($stat.$) \pm 1.7($system.$))\times 10^{-3}$,\\
The statistical error is calculated taking into account the background.
The corresponding constraint is $|A_{\xi}| < 5.44\times 10^{-3} (90\%~ CL) $.

\addtocounter{table}{-1}
A comparison with the result of previous
experiment is given in the table \ref{tab:t1}
for the cuts
$E^{*}_{\gamma} > 10$ MeV, $0.6<cos\Theta^{*}_{e\gamma}<0.9$,
which were used in \cite{e8}.
\begin{table}[ht]
\begin{center}
\begin{tabular}{lrr}
\hline
$A_{\xi}$ & $N_{ev}$ & experiment \\
\hline
$ -0.007 \pm 0.008 \pm0.002$ & 19295 & this experiment \\
$ -0.015\pm0.021 $& 1456 & ISTRA$+$ \cite{e8} \\
\hline
\end{tabular}
\end{center}
\caption{\em Comparison of experimental results\label{tab:t1}}
\end{table}

For the cuts $E^{*}_{\gamma} > 30$ MeV and $\Theta^{*}_{e\gamma}> 20^{\circ}$,
used in theoretical papers \cite{tr2,v13}, the following result is obtained:

$A_{\xi} =(+4.4 \pm 7.9($stat.$) \pm1.9($syst.$))\times10^{-3}$

Let's take a closer look at the assessment of the systematic errors for the case
$E^{*}_{\gamma} > 10 $MeV and $\Theta^{*}_{e\gamma}> 10^{\circ}$.
The systematic  due to selections is determined by varying
each of them and is given in the table \ref{tab:t2}.
Additional systematics occurs from the uncertainty of zero of the $\xi_{\pi e\gamma}$ scale
due to measurement errors.
The evaluation of this contribution gives $\pm$0.00065.

\begin{table}[ht]
\begin{center}
\begin{tabular}[width=0.40\textwidth]{rrrrrrr}
\hline
$N_{cut}$ & 1 & 2 & 3 & 4 & 5 & 6 \\
\hline
$\Delta\cdot10^{3} $& 0.03 & 0.04 & 0.04 & 0.03 & 0.96 & 0.70 \\
\hline
\end{tabular}
\end{center}
\caption{\em The contribution of the variation of each of the cuts to the systematic error\label{tab:t2}
for $E^{*}_{\gamma} > 10$ MeV and $\Theta^{*}_{e\gamma}>10^{\circ}$.}
\end{table}

The estimate of a false asymmetry in the   $\xi_{\pi e\gamma}$ distribution due to the acceptance of the setup,
the efficiency of the reconstruction and selections was carried out using the signal MC,
in which there is no $CP$ violation.
The measured difference between the original (0) and reconstructed value of  the $A_{\xi}$  equals to:
$\Delta_{A} = 0.0012\pm0.0011$ , that is, there is no significant effect.
The error of the estimate is added to the systematics.

The systematics related to the models
used in MC is also investigated. In signal MC $O(p^{4})$
approximation  of ChPT is replaced by
$O(p^{2})$. This gives a negligible effect.

\section{Summary}

The paper continues a study of the $K^+\rightarrow\pi^{0} e^{+}\nu_{e}\gamma$  decay
on statistics of
~$10^{5}$ events three times higher than the one used in \cite{e2}. A search
is performed for $T(CP)$-odd effects in this decay, which could manifest themselves in a non-zero
value of the asymmetry $A_{\xi}$ (2) of the $\xi$ (1) distribution. As a result, 
the values of $A_{\xi}$ for three regions in the photon energy and emission angle
 in the kaon rest frame are obtained: :

$A_{\xi} =(+0.1 \pm 3.9($stat.$) \pm 1.7($syst.$))\times10^{-3}$,
\hspace*{5mm} $|A_{\xi}| < 5.44\times10^{-3} (90\% ~CL)$\\
for $E^{*}_{\gamma} > 10$ MeV and $\Theta^{*}_{e\gamma}> 10^{\circ}$.

$A_{\xi} =(-7.0\pm 8.1($stat.$) \pm $1.5$($syst.$))\times10^{-3}$,
\hspace*{5mm}$|A_{\xi}| < 1.05\times10^{-2} (90\% ~CL)$\\
for $E^{*}_{\gamma} > 10$ MeV, $0.6<cos\Theta^{*}_{e\gamma}<0.9$.

$A_{\xi} =(+4.4 \pm 7.9($stat.$) \pm 1.9($syst.$))\times10^{-3}$, 
\hspace*{5mm} $|A_{\xi}| < 1.04\times10^{-2} (90\% ~CL)$\\
for $E^{*}_{\gamma} > 30$ MeV and $\Theta^{*}_{e\gamma}> 20^{\circ}$.

Within extensions of SM the non-zero asymmetry in the decay
 occur in the vector and axial vector theories, which, in the most general
form, can be described by the matrix element (1), in which the constants $g_{A}$ and $g_{V}$ are complex.
In paper\cite{tr2} in the framework of ChPT $O(p^{4})$ 
the following estimate is obtained:
$A_{\xi}=Im(g_{A}+g_{V}) \times3\cdot10^{-3}$
for $E^{*}_{\gamma}> 30$ MeV and $\Theta^{*}_{e\gamma} > 20^{\circ}$. From here and from our
result we can get the constraint $Im(g_{A}+g_{V}) <3.5 (90\% ~CL)$. This result may
be possibly improved by selecting an optimal region in ($E^{*}_{\gamma}, \Theta^{*}_{e\gamma}$).
For more specific variant of SM-extension, for example, those considered in \cite{t3}
, the estimate $|A_{\xi}| <0.8\times10^{-4}$ was obtained in \cite{tr2}.

The work was carried out with the support of the RSCF grant No. 22-12-0051 .


\begin{thebibliography}{99}

\bibitem{e2}
A.~Yu.~Polyarush, S.A. Akimenko, A.V. Artamonov et al. (OKA),
Eur. Phys. J. C {\bf 81}, no.2, 161 (2021).

\bibitem{v1}
N. Cabibbo  Phys.Rev.Lett {\bf 10}, 531 (1963).

\bibitem{v2}
Kobayashi M., Maskawa T. Progr.Theor. Phys. {\bf49}, 652 (1973).
\bibitem{v3}

C. Jarlskog Z.Phys. {\bf C29}, 491 (1985).
\bibitem{v4}

G.F.Farrar and M.E.Shaposhnikov, Phys.Rev.Lett {\bf 70}, 2833(1993), [Erratum ibid 
{\bf 71}(1993)210],hep-ph/9305274.

\bibitem{v42}
P. Het and E. Sather, Phys.Rev {\bf D51}(1995)379, hep-ph/9404302.

\bibitem{v43}
M.Carena, M.Quiros and C.E.Wagner, Phys.Lett. {\bf B380}, 81 (1996),  hep-ph/9303420.

\bibitem{v6}
J.F.Donaghue, B.Holstein Phys.Lett.{\bf B113}, 382 (1982).

\bibitem{v61}
L.Wolfenstein Phys. Rev. {\bf D29}, 2130 (1984).

\bibitem{v62}
G.Barenboim, J. Bernabeu, J. Prades and M. Raidal Phys. Rev. {\bf D55}, 4213 (1997).

\bibitem{v7}
M.Koboyashi,T.T.Lin, Y.Okada Progr. Theor. Phys. {\bf 95}, 361 (1996).

\bibitem{v71}
R.Garisto, G.Kane Phys. Rev. {\bf D44}, 2038 (1991).

\bibitem{v8}
G.Belanger, C.Q.Geng Phys. Rev.{\bf D44}, 2789 (1991).

\bibitem{v9}
Y. Grossman, Y. Nir Phys.Lett. {\bf B313}, 126 (1993). 

\bibitem{v10}
G.H. Wu and John N. Ng Phys.Lett {\bf B392}, 93 (1997).

\bibitem{v11}
M. Abe,  M. Aliev, V. Anisimovsky, et al.,(KEK-E246 Collaboration) Phys. Rev. {\bf D73}, 072005 (2006).

\bibitem{v111}
V.V. Anisimovsky, A.N. Khotjantsev, A.P. Ivashkin, Phys. Lett. {\bf B562}, 166 (2003).

\bibitem{tr2} 
V. V. Braguta, A. A. Likhoded, A. E. Chalov, Phys. Rev. {\bf D68}, 094008 (2003).


\bibitem{v12}
J.Gevas, J.Iliopolus, J.Kaplan Phys. Lett.{\bf 20}, 432 (1966). 

 \bibitem{v5}
V.V.Braguta,A.A.Likhoded, A.E.Chalov, Phys. Rev. {\bf D65}, 054038 (2002).


\bibitem{v13}
I.B. Khriplovich, A.S. Rudenko, Phys. Atom. Nucl. {\bf 74}, 1214 (2011).

\bibitem{e3}
V.~S.~Burtovoy, S.A. Akimenko, A.V. Artamonov et al. (OKA) J. Exp. Theor. Phys. {\bf 131}, no.6, 928 (2020).

\bibitem{e4}
V.~I.~Kravtsov S.A. Akimenko, A.V. Artamonov et al. (OKA),
Eur. Phys. J. C {\bf 79}, no.7, 635 (2019).

\bibitem{e9}
R. Brun, F. Bruyant, M. Maire. et al. CERN-DD/EE/84-1, Geneva, 1987.

\bibitem{e8}
S.A. Akimenko, V.N.Bolotov, G.I. Britvich   et al. (ISTRA+), YaF {\bf 70}, 1 (2007), Phys. Atom. Nucl. {\bf 70}, 29 (2007).

\bibitem{t3} 
J.C. Pati and A. Salam, Phys.Rev. {\bf D10}, 275 (1975).

\end{thebibliography}
\end{document}